\documentclass[submitting]{nst}

\usepackage{subfigure,dcolumn}
\usepackage{epstopdf}
\usepackage{mhchem}
\usepackage{upgreek}

\begin{document}

\title{Number-of-constituent-quark scaling of elliptic flow: a quantitative study}
\thanks{This work was supported partly by the National Natural Science Foundation of China (Nos. 11905120 and 11947416), the U.S. Department of Energy (No. DE-FG03-93ER40773), and NNSA Grant No. DENA0003841 (CENTAUR).}

\author{Meng Wang}
\affiliation{School of Physics \& Information Technology, Shaanxi Normal University, Xi’an 710119, China}

\author{Jun-Qi Tao}
\affiliation{School of Physics \& Information Technology, Shaanxi Normal University, Xi’an 710119, China}

\author{Hua Zheng}
\email[Corresponding author, ]{zhengh@snnu.edu.cn}
\affiliation{School of Physics \& Information Technology, Shaanxi Normal University, Xi’an 710119, China}

\author{Wen-Chao Zhang}
\affiliation{School of Physics \& Information Technology, Shaanxi Normal University, Xi’an 710119, China}

\author{Li-Lin Zhu}
\affiliation{Department of Physics, Sichuan University, Chengdu 610064, China}

\author{Aldo Bonasera}
\affiliation{Cyclotron Institute, Texas A\&M University, College Station, TX 77843, USA}
\affiliation{Laboratori Nazionali del Sud, INFN, via Santa Sofia, 62, 95123 Catania, Italy}

\begin{abstract}
The number-of-constituent-quark (NCQ) scaling behavior of the elliptic flow of identified particles produced in A+A collisions is studied quantitatively using an empirical function that fits the experimental $v_2$ data available from the RHIC and LHC. The most common approach for NCQ scaling involves (1) doing a scaling of the experimental $v_2$ data of an identified particle with its NCQ, (2) doing the same to its transverse momentum or energy, then (3) combining all the scaled data and identifying the NCQ behavior by intuitively looking (since the measured experimental data are discrete). We define two variables $(d_1, d_2)$ to describe NCQ scaling quantitatively and simultaneously, and identify the approximate region where the NCQ scaling holds. This approach could be applied to study NCQ or other scaling phenomena in future experiments.
\end{abstract}


\maketitle

\section{Introduction}\label{sec.I}

The main goal of producing relativistic nucleus-nucleus collisions at the RHIC and LHC is to create a deconfined quark and gluon plasma (QGP), a new state of matter that forms at a high density and temperature, as predicted by quantum chromodynamics (QCD), and to understand its properties \cite{ref1,ref2,kh,tat}. The azimuthal anisotropies of the particles produced in the collisions have proven to be a powerful probe for investigating QGP characteristics and hadron structure \cite{ref4, ref6, ref9, ref11, ref32, ref33, ref34, ref35, ref36, ref37,Lin:2021mdn,Tang:2020ame}. Therefore, they have been extensively measured experimentally and studied both experimentally and theoretically. These anisotropies can be quantified in terms of the coefficients $v_n$ in the Fourier-series expansion of the particle distributions with respect to the reaction plane (RP), defined by the beam axis and the impact parameter, which is determined on an event-by-event basis \cite{ref10, ref17, ref18, ref28, ref29}:
\begin{equation}
E\frac{\mathrm{d}^{3}N}{\mathrm{d}p^{3}}=\frac{\mathrm{d}^{2}N}{2\pi p_\text{T}\mathrm{d}p_\text{T}\mathrm{d}y}\left(1+\sum _{n=1}^{\infty}2v_{n}\cos[n(\varphi-\Psi _\text{RP})]\right),
\end{equation}
where $\varphi$ and $\Psi_\text{RP}$ are the azimuthal angles of the particle of interest and of the reaction plane, respectively. The Fourier coefficient is given by
\begin{equation}
v_n=\langle\cos[n(\varphi-\Psi _\text{RP})]\rangle.
\end{equation}

In practice, several methods have been proposed for analyzing the azimuthal anisotropies of the final particles. These include the event-plane method, the $\eta$-subevent method, the cumulant method, and the Lee-Yang-Zero (LYZ) method \cite{ref3,ref4, ref13, ref21}. Their purpose is to mitigate the non-flow contributions and the flow fluctuations. In the past two decades, the direct flow ($v_1$) and  elliptic flow ($v_2$) have been measured experimentally at both the RHIC and LHC. One remarkable finding of the flow investigations conducted at the RHIC (and confirmed at the LHC) is the scaling of $v_2$ with the number of constituent quarks in a hadron, in cases where many particle species produced in high-energy nucleus-nucleus collisions in the low-transverse-momentum region \cite{ref4,ref6,ref9, ref32, ref11,ref15,ref28}. Furthermore, at both the RHIC and LHC, this scaling is seen to improve when expressed as a function of the scaled hadron transverse energy \cite{ref4,ref6,ref9,ref11,ref15}. This may be a consequence of energy conservation, as discussed in Ref. \cite{ref33} based on the Boltzmann equation. This observation is consistent with the standard model, which considers quarks as the basic building blocks of all matter, and with quark coalescence as the hadronization mechanism \cite{ref32, ref33, ref34, ref35, ref36, ref37, ref7,ref10,ref23,ref24,ref29,  Das:1977cp, Hwa:2002zu, Hwa:2002tu, Hwa:2006vb, Hwa:2011bw, Zhu:2013cza, Chen:2018tnh}. This provides indirect evidence that a QGP is formed in high-energy nucleus-nucleus collisions. 

Usually in experimental investigations of number-of-constituent-quark (NCQ) scaling phenomena, both the experimental $v_2$ data and the transverse momentum or energy of an identified particle species are scaled with its number of constituent quarks. Then, the scaled $v_2$ is plotted as a function of the scaled transverse momentum or energy. NCQ scaling is then identified by eye because the experimental data measured are discrete. Several attempts have been made to quantitatively elucidate the NCQ scaling \cite{ref4, ref9, ref12,ref13}. Such studies adopt a polynomial function to fit the scaled $v_2$ as a function of the scaled transverse momentum or energy for a chosen particle species. (The chosen order of the polynomial, up to the seventh, depends on the particle of interest, the collision energy, and the fitting range of the scaled transverse momentum or energy.) This polynomial defines a baseline from which to calculate the deviation of the scaled $v_2$ for the other particle species. However, such a polynomial function is suboptimal because it may oscillate, and its behavior beyond the data range is determined by the sign of the coefficient of the largest-order term. Other forms of empirical functions have also been proposed to investigate the NCQ scaling by fitting the $v_2$ data for mesons and baryons simultaneously \cite{ref10,ref14, ref15}. These are discussed in the next section. Considering the fact that the $p_\text{T}$ bins differ for mesons and baryons in experimental measurements, we here propose, as a plausible quantitative approach, to search for an empirical function capable of fitting all the experimental $v_2$ data and to conduct the investigation based on this analytical function. 

The remainder of this paper is organized as follows. Section 2 briefly introduces two variables $(d_1, d_2)$ to quantify the NCQ scaling of the identified particles simultaneously. The empirical functions quoted in the literature and our proposed empirical function for fitting the experimental $v_2$ data are also introduced. Section 3 shows the fit results of $v_2$ for different particle species from the beam energy scan (BES) program at the RHIC to LHC. The NCQ scaling is shown quantitatively to be a function of both the scaled transverse momentum and the scaled transverse energy. Section 4 closes with concluding remarks.

\begin{figure*}[t!]
    \centering
    \begin{tabular}{cc}
     \includegraphics[width=.5\textwidth]{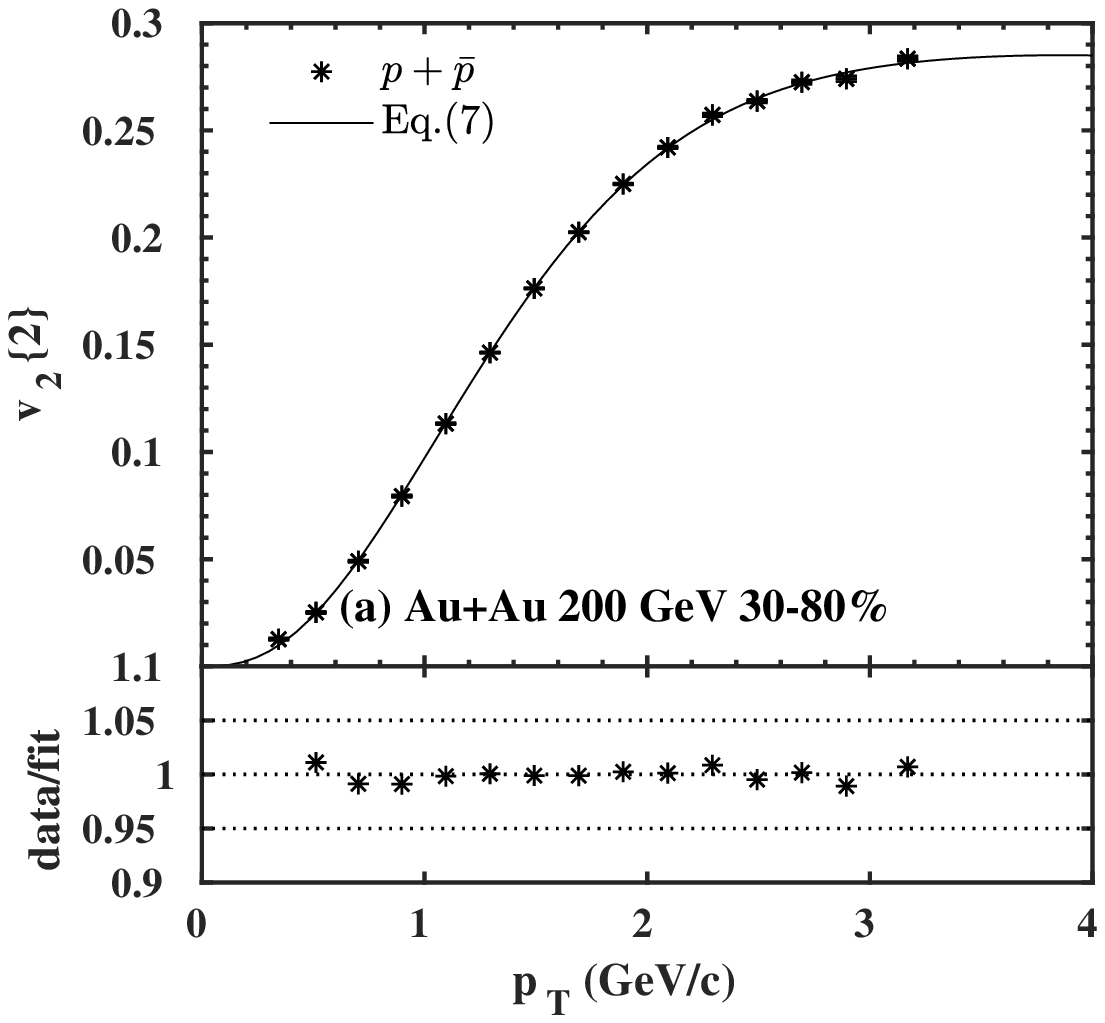} \includegraphics[width=.5\textwidth]{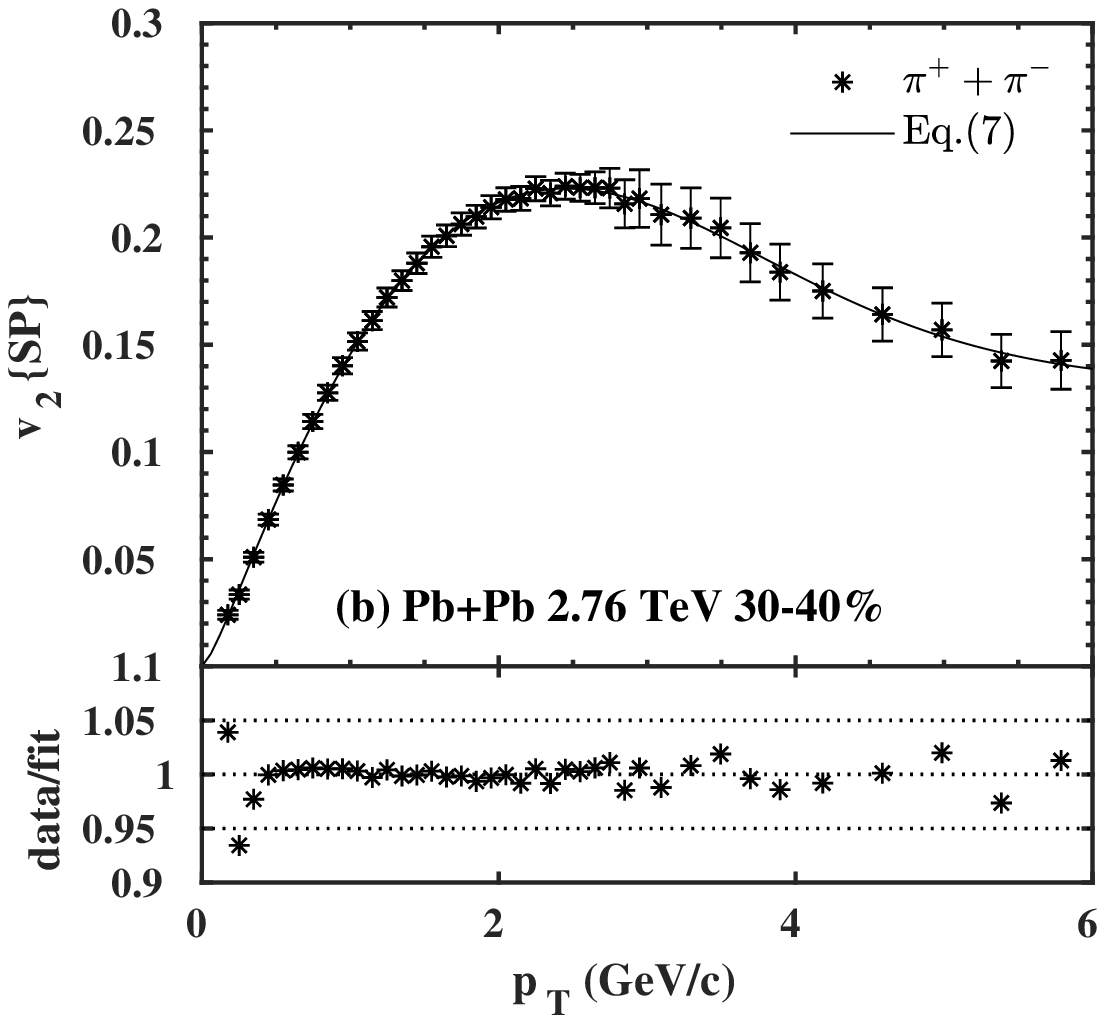}
     \end{tabular}
\caption{Dependence of $v_2$ on $p_\text{T}$ for (a) protons in Au+Au collisions at $\sqrt{s_\text{NN}}=200$ GeV with a centrality of 30--80\%, and (b) for pions in Pb+Pb collisions at $\sqrt{s_\text{NN}}=2.76$ TeV with a centrality of 30--40\%. The data are obtained from Refs. \cite{ref9, ref11}. The methods used to extract $v_2$ are labeled. The curves are the fit results for the $v_2$ data using  Eq. (\ref{eqour}). The data/fit are shown in the bottom panels, respectively.}    
\label{figure1}
\end{figure*}

\section{Method and empirical functions}\label{sec.2}

Assuming that we know the information at any point of the scaled $v_2$ and the corresponding scaled transverse momentum or energy for each particle species, we can define two variables that simultaneously and quantitatively characterize the quality of the NCQ scaling. 

The first variable is the deviation between the scaled $v_2$ for one particle species and the average of all the scaled $v_2$ values for the particle species of interest in the NCQ scaling at each scaled transverse momentum or energy. It can thus be defined as
\begin{equation}
d_1(\tilde p_\text{T}) =\left |\tilde v_2^i(\tilde p_\text{T}) - \overline{\tilde v_2(\tilde p_\text{T})}\right|, \label{eqd1}
\end{equation}
where $\tilde p_\text{T}=\frac{p_\text{T}^i}{n_i}$ is the scaled transverse momentum, and $n_i=2$ for the mesons and 3 for baryons. We denote as $\tilde v_2^i(\tilde p_\text{T})$ the scaled $v_2$ for particle species $i$, and as $\overline{\tilde v_2(\tilde p_\text{T})}$ the average of all the scaled $v_2$ of the particle species of interest. Equation (\ref{eqd1}) assumes that $\overline{\tilde v_2(\tilde p_\text{T})}$ is the real NCQ scaling curve. If the NCQ scaling is perfect, then $d_1(\tilde p_\text{T}) =0$ for each $\tilde p_\text{T}$. Therefore, the deviation of $d_1(\tilde p_\text{T})$ from 0 can characterize the extent by which the NCQ scaling is violated.

The second variable is the difference between the maximum of $\tilde v_2^i(\tilde p_\text{T})$ and the minimum of $\tilde v_2^j(\tilde p_\text{T})$ among the particle species of interest at a given $\tilde p_\text{T}$:
\begin{equation}
d_2(\tilde p_\text{T})=\textrm{max}\{\tilde v_2^i(\tilde p_\text{T})\}-\textrm{min}\{\tilde v_2^j(\tilde p_\text{T})\}. \label{eqd2}
\end{equation}
Notably, the particle species $i,j$ can differ when $\tilde p_\text{T}$ changes and $d_2(\tilde p_\text{T})$ is positively defined. Similar to $d_1(\tilde p_\text{T})$, for ideal NCQ scaling, $d_2(\tilde p_\text{T})$ should equal zero for each $\tilde p_\text{T}$. Again, a non-zero $d_2(\tilde p_\text{T})$ can be used to characterize the extent of the NCQ scaling violation.

When we apply the scaled transverse energy $\tilde E_\text{T}=\frac{E_\text{T}^i}{n_i}$, where $E^i_\text{T}=\sqrt{{m^i_0}^2+{p^i_\text{T}}^2}-m^i_0$ and $m^i_0$ is the rest mass of particle $i$, the definitions of the two variables in Eqs. (\ref{eqd1}, \ref{eqd2}) can also be used by replacing $\tilde p_\text{T}$ with $\tilde E_\text{T}$. 

The discreteness of the experimental data precludes their simultaneously satisfying the above assumption for both the scaled transverse momentum and energy. This is because the particle masses differ and the conversion between $p_\text{T}$ and $E_\text{T}$ is nonlinear. Therefore, to bridge the gap between the data and our requirements, we utilize an empirical function that provides a good fit to the experimental data. 

Several empirical functions have been proposed in the literature to fit $v_2$ data. The most popular one adopted in the experimental papers is a simple polynomial function. As argued above, it has defects that also become apparent when choosing polynomial orders between 3 and 7 \cite{ref4, ref9,ref12,ref13}. The second best-known empirical function was proposed by Dong {\it et al.} \cite{ref10} when NCQ scaling was discovered at the RHIC:
\begin{equation}
f_{v_{2}}(p_\text{T}, n)=\frac{an}{1+\exp[-(p_\text{T}/n-b)/c]}-dn.  \label{eqxu1} 
\end{equation}
Equation (\ref{eqxu1}) has four fitting parameters $a,b,c$, and $d$, with $n$ being the NCQ in a particle species. Clearly, $f_{v_{2}}(p_\text{T}, n)$ becomes a constant at high $p_\text{T}$, contradicting recent experimental data measured at high $p_\text{T}$ \cite{ref3, ref4, ref9, ref11, ref15,ref17, ref18, ref24,ref28, ref31}. However, it was good to use at the early times because only low $p_\text{T}$ region data were measured back to that time.  Equation (\ref{eqxu1}) appears not to include the origin, as required by the definition of $v_2$. 

To overcome these limitations, another empirical function was proposed in Ref. \cite{ref14}:
\begin{equation}
v_{2}(p_\text{T}, n)=\frac{p_{0} n}{1+\exp(\frac{p_1-p_\text{T}/n}{p_2})}-\frac{p_{0} n}{1+\exp(\frac{p_1}{p_2})}-p_{3}n p_\text{T}, \label{eqxu2} 
\end{equation}
where $p_0$, $p_1$, $p_2$, and $p_3$ are free parameters. This equation  is satisfied at the origin and is not constant at high $p_\text{T}$. As shown in the next section, it can fit most of the available experimental $v_2$ data but not for large $p_\text{T}$. We therefore do not adopt Eq. ~ (\ref{eqxu2}) in our study. 

As an alternative, we propose a new empirical function:
\begin{equation}
v_{2}(p_\text{T})=\frac{ap_\text{T}^b}{1+\exp\left[dp_\text{T}^c+\tanh\left(ep_\text{T}^f\right)\right]}, \label{eqour} 
\end{equation}
with free parameters $a, b, c, d, e$, and $f$. We emphasize that the experimental $v_2$ data are well fitted by Eq. (\ref{eqour}), as discussed in the next section.

\begin{figure*}[t!]
\centering
    \begin{tabular}{c}
  \includegraphics[scale=0.35]{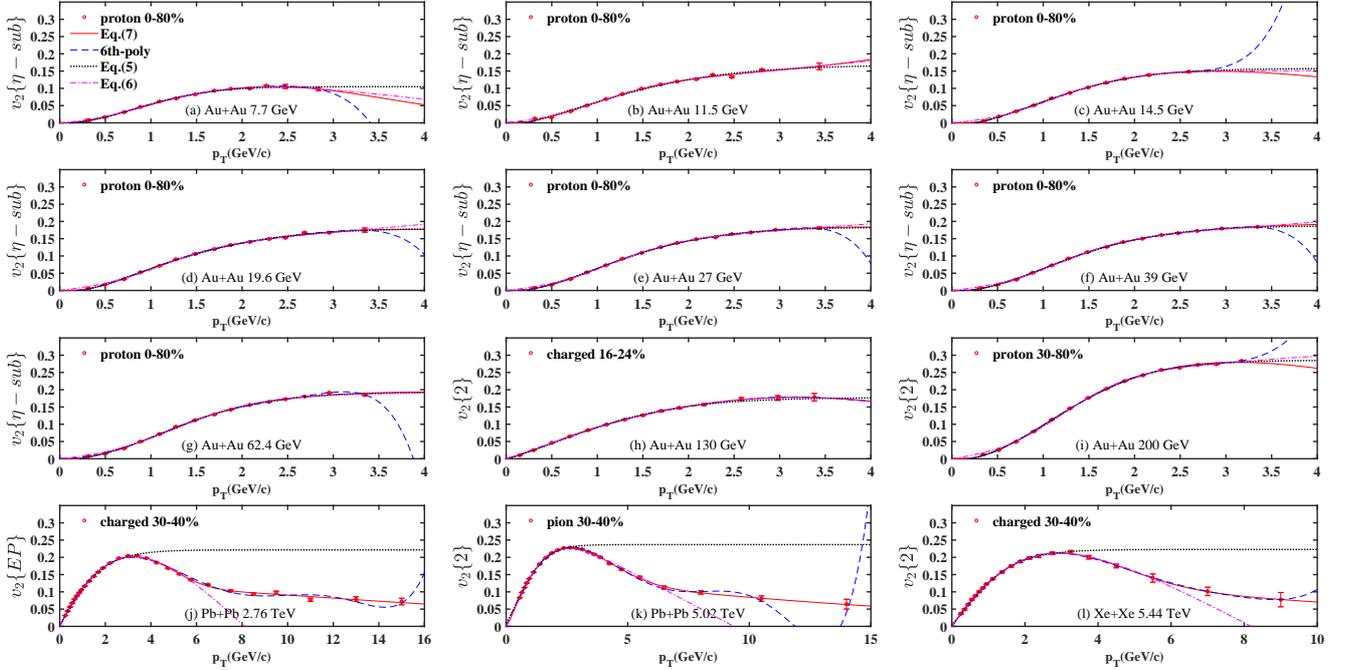}
\end{tabular}
\caption{(Color online) Dependence of $v_2$ on $p_\text{T}$ for protons or charged particles produced in Au+Au collisions at the RHIC, and for pions or charged particles produced in Pb+Pb or Xe+Xe collisions at the LHC. The data are obtained from Refs. \cite{ref9,ref15,ref18,ref21,ref28,ref27}. The labels indicate the methods used to extract $v_2$. The curves are the fit results to the $v_2$ data using Eqs.(\ref{eqxu1}, \ref{eqxu2}, and \ref{eqour}) and the 6th-order polynomial, as indicated in the legend.} 
\label{figure2}  
\end{figure*}

\begin{figure*}[t!]
    \centering
     \includegraphics[width=.45\textwidth]{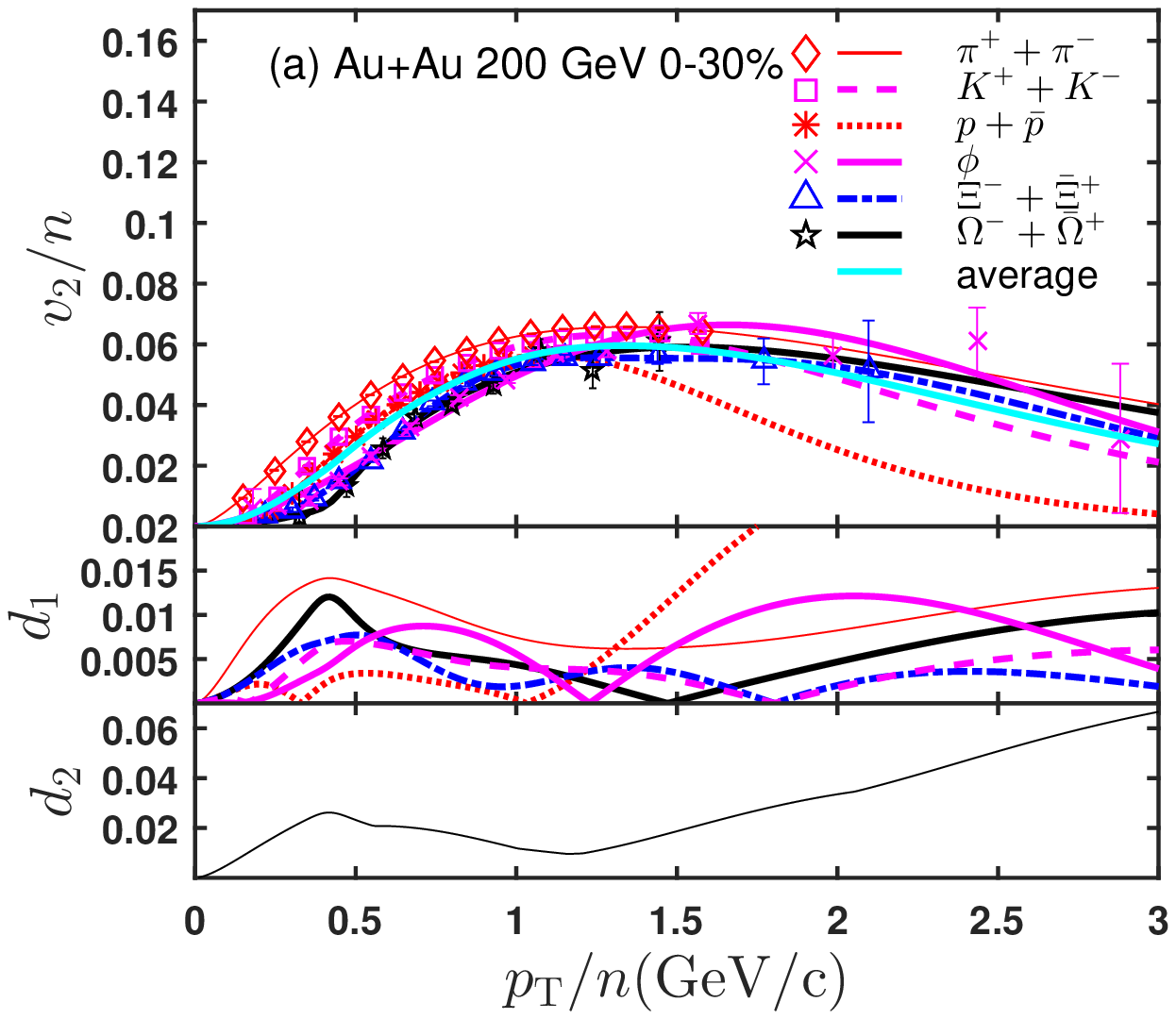}\includegraphics[width=.45\textwidth]{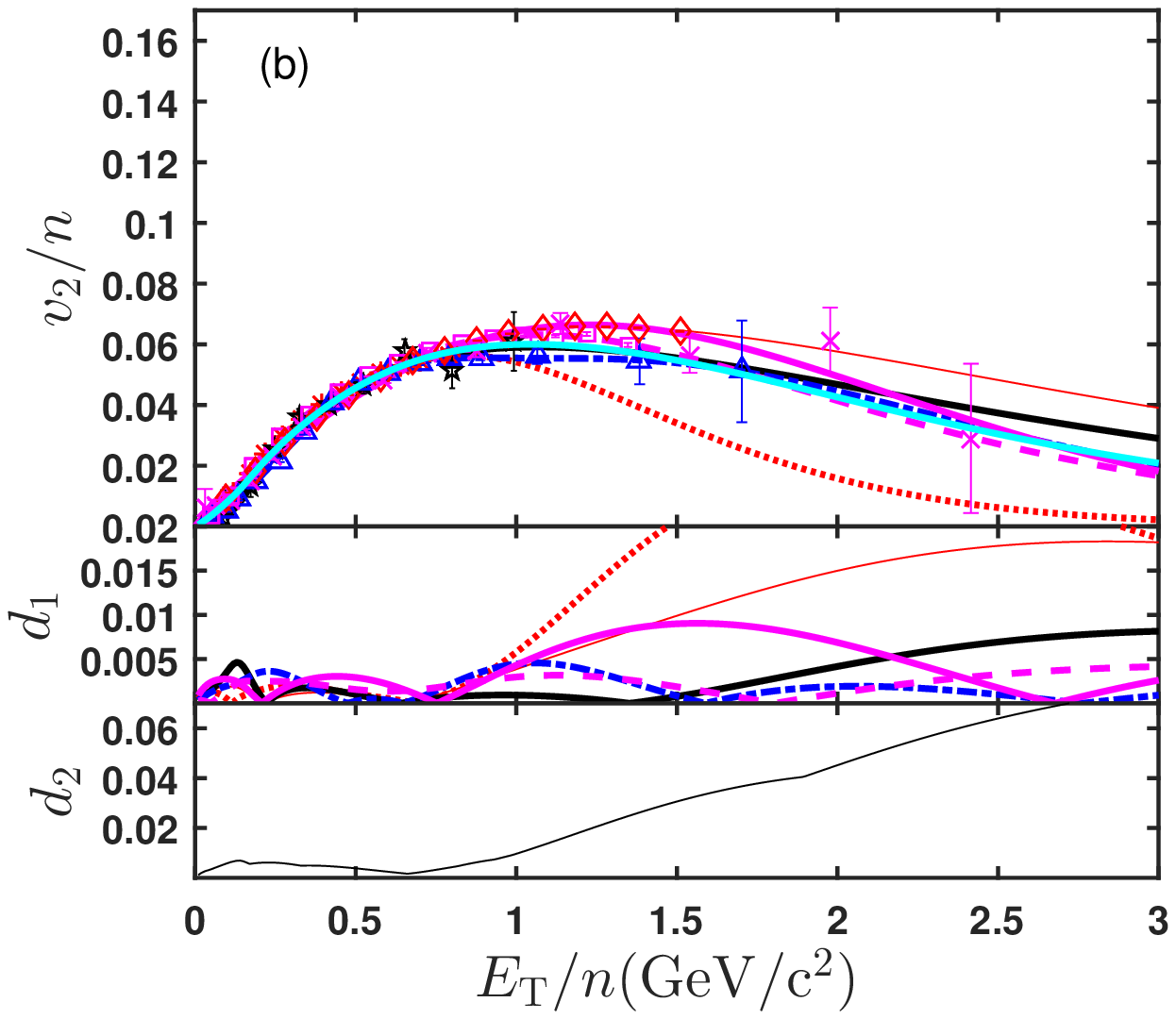}
\caption{(Color online) Scaled elliptic flow versus scaled transverse momentum (left panel) and energy (right panel) for the six identified particles (indicated in the legend) produced in Au+Au collisions at $\sqrt{s_\text{NN}}=200$ GeV with centrality 0-30\%. The curves for the fit results obtained using Eq. (\ref{eqour}) and their averages are plotted. The two variables ($d_1, d_2$) are also plotted as functions of the scaled transverse momentum and energy, with appropriate scales for showing the details where NCQ scaling holds. The data are obtained from Ref. \cite{ref9}.}
\label{figure3}  
\end{figure*}

\section{Results}\label{sec.3}

We test the empirical function in Eq. (\ref{eqour}) by fitting the $v_2$ data for the available identified or charged particles, ignoring data that either display large fluctuations or are scant for A+A collisions from the RHIC and LHC (i.e., from $\sqrt{s_\text{NN}}=$7.7 GeV to 5.44 TeV). This is because the $v_2$ data with large fluctuations do not have sufficient statistics, and those with only few data points can not constrain the free parameters in Eq. (\ref{eqour}). Neither case can serve the purpose of this study. The long-running and successful high-energy heavy-ion collision program still provides a wealth of $v_2$ data suitable for our investigation. Notably, Eq. (\ref{eqour}) is positively defined. On the rare occasions where, for some particle, $v_2$ is negative when $p_\text{T}$ is close to zero \cite{ref24, ref25, ref26}, we ignore that data point in the fitting.  Figures \ref{figure1} and \ref{figure2} show our fit results for Eq. (\ref{eqour}) to $v_2$ data for the selected particles in different collision systems and at different collision energies. 

Figure \ref{figure1} shows examples of fit results for (a) the $v_2$ data of $p+\bar{p}$ derived from Au+Au collisions at $\sqrt{s_\text{NN}}=200$ GeV with centrality 30--80\%, and (b) the $v_2$ data of $\pi^++\pi^-$ derived from Pb+Pb collisions at $\sqrt{s_\text{NN}}=2.76$ TeV with centrality 30--40\%. The ratio of the data to the fit values provides a visual assessment of the fit quality, reflecting the discrepancy between the central values of the data and the fit results shown in the bottom panel for each subfigure. For both cases, the central values of the $v_2$ data deviate from Eq. (\ref{eqour}) within 5\%, except for one data point from Au+Au collisions at $p_\text{T}$ close to 0. The very low value at that point makes the data/fit ratio very sensitive to the fit. 

Figure \ref{figure2} shows the fit results with Eq. (\ref{eqour}) to the $v_2$ data for different centralities and collision energies. Subplots (a)-(g) show $v_2$ for the elliptic flow of protons in Au+Au collisions at $\sqrt{s_\text{NN}}= $ 7.7 GeV to 62.4 GeV with centrality $0-80\%$.  Subplots (h) and (i) plot the elliptic flow of charged particles in Au+Au collisions at $ \sqrt{s_\text{NN}}=$ 130 GeV with centrality $16-24\%$ and of protons in Au+Au collisions at $\sqrt{s_\text{NN}}= $200 GeV with centrality $30-80\%$. Subplots (j) to (l) give the elliptic flow of charged particles (pions) in Pb+Pb collisions at $ \sqrt{s_\text{NN}}=2.76, 5.02 $ TeV and in Xe+Xe collisions at $ \sqrt{s_\text{NN}}=5.44 $ TeV with centrality $30-40\%$ \cite{ref9,ref15,ref18,ref21,ref28}. Good fits are also obtained for the other tested cases. Exhaustive testing suggested that Eq. (\ref{eqour}) can fit the $v_2$ data extracted from different approaches, for its central values well over a wide range of $p_\text{T}$, which definitely covers the $p_\text{T}$ range where the NCQ scaling holds. 

\begin{figure*}[t!]
    \centering
     \includegraphics[width=.45\textwidth]{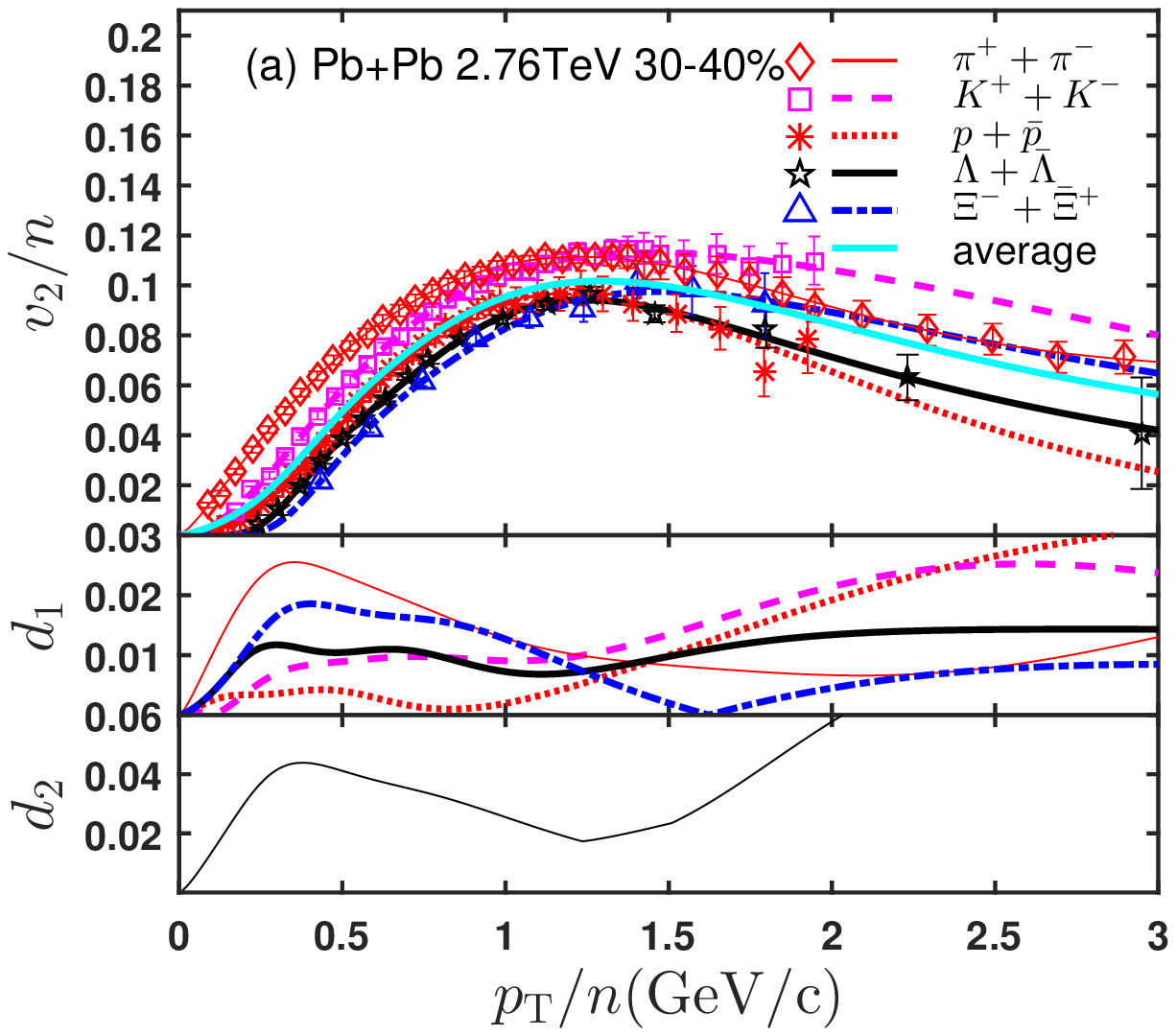}  \includegraphics[width=.45\textwidth]{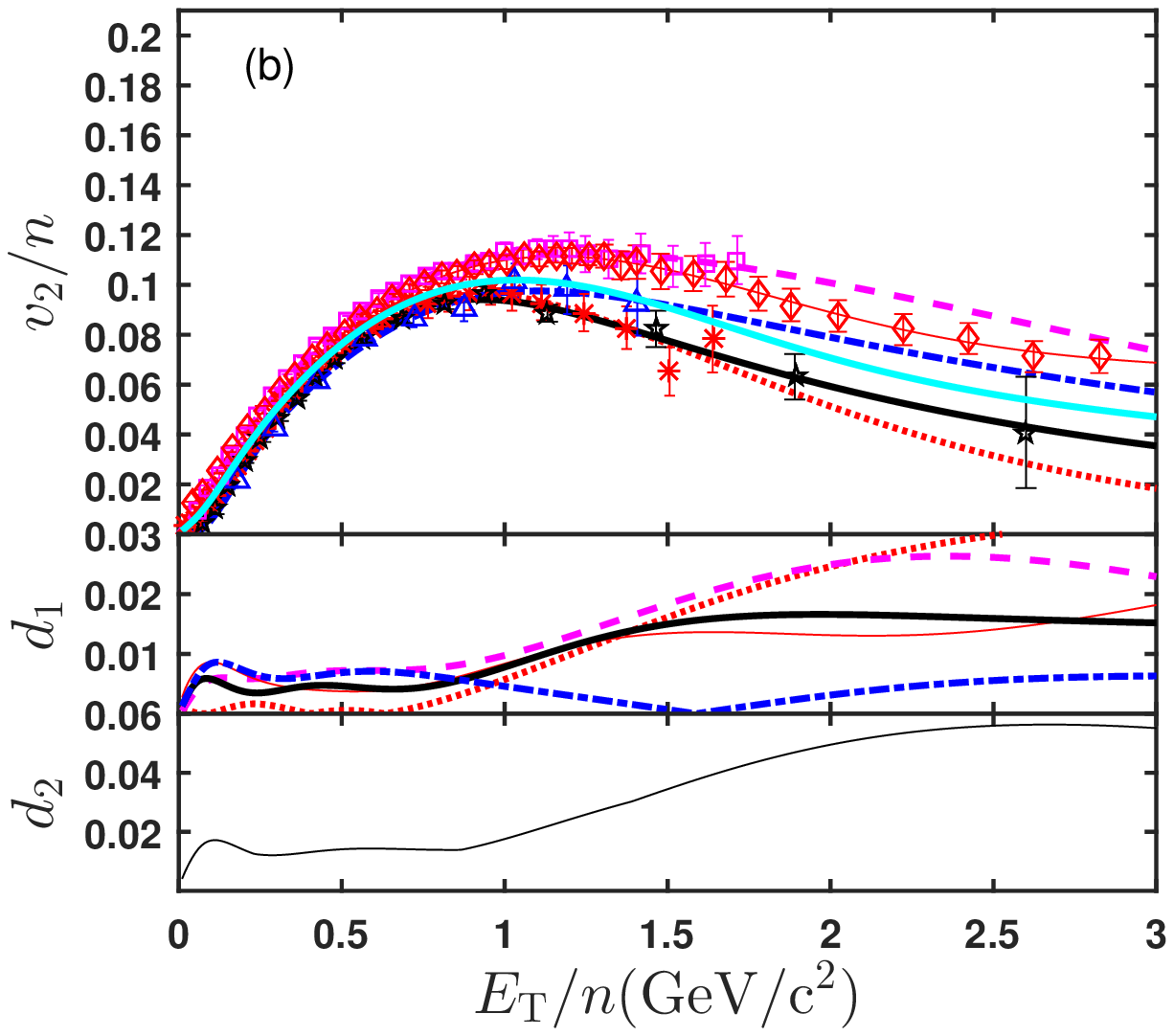}
\caption{(Color online) Similar results to Fig. \ref{figure3} but for Pb+Pb collisions at $\sqrt{s_\text{NN}}= $2.76 TeV with $30-40\%$ centrality. The data are obtained from Ref. \cite{ref11}.}
\label{figure4}  
\end{figure*}  

The fit results from the functions frequently adopted in the experimental papers, i.e., the polynomial function, Eqs. (\ref{eqxu1}) and (\ref{eqxu2}), are also shown. For the polynomail function, we choose a 6th-order polynomial as an example. The fitted curve at high $p_\text{T}$ clearly either increases or decreases depending on whether the sign of the coefficient of the highest order term is, respectively, positive or negative. Extrapolating the results based on this polynomial function is surely unreliable. Oscillations also appear when the fitting range is large. The empirical function in Eq. (\ref{eqxu1}) can only fit the $v_2$ data for low and intermediate $p_\text{T}$ values, where $v_2$ reaches its maximum because Eq. (\ref{eqxu1}) is constant at large $p_\text{T}$ by definition. Equation (\ref{eqxu1}) is not guaranteed to equal zero at $p_\text{T}=0$ GeV/$c$. Equation (\ref{eqxu2}) is an improvement of Eq. (\ref{eqxu1}) and can fit almost all of the $v_2$ data presented. We therefore adopted the empirical function in Eq.~(\ref{eqour}). There is a noticeable limitation for the extrapolation from the fit, which depends on how well the fitting function is constrained beyond the data points. This issue, common to all fitting functions, has consequences for our analysis, as discussed below. Therefore, the two variables $d_1$ and $d_2$, defined in Sect. 2, are utilized to mitigate the problem caused by extrapolation beyond the data points. Fortunately, there is no need to extrapolate in the NCQ scaling region, and our conclusion is not affected.

We can now quantitatively investigate the NCQ scaling of elliptic flow as a function of the scaled transverse momentum (or energy) of the identified particles produced in A+A collisions at the RHIC and LHC. Figures \ref{figure3} and \ref{figure4} show, respectively, the NCQ scaling of the elliptic flow $v_2$ of six identified particles, i.e., $\pi^+ + \pi^-$, $K^++K^-$, $p+\bar{p}$, $\phi$, $\Xi^-+\bar{\Xi}^+$ and $\Omega^-+\bar{\Omega}^+$ from Au+Au collisions at 200 GeV with centrality $0-30\%$ at the RHIC; and the elliptic flow $v_2$ of five identified particles, i.e., $\pi^+ + \pi^-$, $K^+ + K^-$, $p+\bar{p}$, $\Lambda+\bar{\Lambda}$ and $\Xi^-+\bar{\Xi}^+$ from Pb+Pb collisions at 2.76 TeV with centrality $30-40\%$ at the LHC. The values of $v_2$ for an identified particle are known to depend on the extraction method \cite{ref3,ref4, ref13, ref21}. This could affect the NCQ scaling when the $v_2$ data for different particle species of interest were extracted using different approaches in experiments. Therefore, we select the $v_2$ of the different identified particles extracted by the same method for one chosen collision system. The fit to the experimental data, obtained using Eq. (\ref{eqour}) and the two variables $(d_1, d_2)$ defined above, versus the scaled transverse momentum (or energy), are also shown. 

Figure \ref{figure3} (a) presents the NCQ scaling of the elliptic flow for the six identified particles at a centrality of 0-30\% from Au+Au collisions at $\sqrt{s_\text{NN}}= $200 GeV versus the scaled transverse momentum. Figure \ref{figure3} (b) shows the scaling versus the scaled transverse energy. The experimental data clearly scale better versus the scaled transverse energy, as discussed in Ref. \cite{ref33}. In Fig. \ref{figure3} (a), the scaled elliptic flow data points are clearly ordered according to mass, except $\phi$, which refers to protons at very low $\tilde{p}_\text{T}$. In other words, lighter particles have a larger scaled $\tilde{v}_2$ at the same scaled transverse momentum in the low $p_\text{T}$ region, which refers to the mass ordering of the scaled elliptic flow. This anomaly of mass ordering between $\phi$ and protons was observed and explained in Ref. \cite{ref9}. It is seen that $\pi$ stays relatively far from the other particles, which was interpreted in Refs. \cite{ref34, ref10} as being due to the resonance decays. When the scaled $\tilde{v}_2$ are plotted against the scaled transverse energy in Fig. \ref{figure3} (b), all the data points map onto a single curve and the mass ordering vanishes at low $\tilde{E}_\text{T}$. These results are clearly illustrated by the two variables $(d_1, d_2)$ in the bottom panels of Fig. \ref{figure3}. We note that the extrapolation from the fitting curve for $p+\bar{p}$ is poor because no data points are available in the high $p_\text{T}$ region and show different behavior from other particles. According to Eq. (\ref{eqd2}), $d_2$ is affected most strongly in the extrapolation region. This effect is mitigated in the case of $d_1$ by an averaging over all the particle species. We also emphasize that the region where NCQ scaling holds is not affected. Comparing the values from  Figs. \ref{figure3} (a) and (b), we see that both $d_1$ and $d_2$ deviate slightly from 0 at $p_\text{T}/n<1.2$ GeV/$c$, which is the crossing point of the scaled $\tilde{v}_2$ for several particle species. On the other hand, $d_1$ and $d_2$ are very close to 0 at $E_\text{T}/n<1.0$ GeV/$c^2$. For the other scaled transverse momentum (or energy) region where the NCQ scaling is completely violated, $d_1$ and $d_2$ increase and deviate from 0. The consistent behavior of $d_1$ and $d_2$ is not surprising because they are both defined to reflect the quality of the NCQ scaling from different perspectives. Our results suggest not only that they provide  consistency between the NCQ scaling and that done intuitively by looking; they also show fine details.

Figures \ref{figure4} (a) and (b) show, respectively, the NCQ scaling of elliptic flow for the five identified particles at centrality 30-40\% from Pb+Pb collisions at $\sqrt{s_\text{NN}}= $2.76 TeV, versus scaled transverse momentum and scaled transverse energy. Similar results to those in Fig. \ref{figure3} are observed generally. The NCQ scaling violation is clearly evidenced because the data are available at high $p_\text{T}$ for all particle species of interest at the LHC. The $d_1$ and $d_2$ variables are much less affected by the extrapolation issue for the same reason. The mass ordering of scaled $v_2$ versus $p_\text{T}/n$ is unquestionable. 
These results reflect the fact that the same matter is created in the relativistic nucleus-nucleus collisions, and that it undergoes similar dynamical processes. We emphasize that the conclusion does not change from Au+Au at RHIC in Fig. \ref{figure3}.

\section{Conclusion}\label{sec.4}
To summarize, we have proposed an empirical function for fitting the elliptic flow $v_2$ data of the identified particles produced in A+A collisions at the RHIC and LHC. We also quantitatively investigate the NCQ scaling of the elliptic flow of the identified particles. This is done by utilizing the analytical empirical function to overcome the challenge posed by the discreteness of the experimental data. Thus, the NCQ scaling cannot be quantitatively investigated simultaneously for the scaled transverse momentum and energy.  Given the issues associated with the extrapolation from the fitting function beyond the measured data region, particularly for Au+Au at the RHIC, two variables $(d_1, d_2)$ are defined to quantify the NCQ scaling simultaneously.  As expected, they not only give consistent results with those obtained by intuitively looking at the data (namely, that the NCQ scaling is better for the scaled transverse energy than the scaled transverse momentum); they also provide fine details of the region where the NCQ holds. This approach can be applied to study other experimental scaling phenomena quantitatively.

\section*{Author contributions}
All authors contributed to the study conception and design. Material preparation, data collection and analysis were performed by Meng Wang, Jun-Qi Tao and Hua Zheng. The first draft of the manuscript was written by Hua Zheng and all authors commented on previous versions of the manuscript. All authors read and approved the final manuscript.

\end{document}